\documentstyle[12pt,epsfig]{article}
\title{On the spectrum of relativistic Schroedinger equation
in finite differences.}
\author{V.A.Berezin, A.M.Boyarsky\\
Institute for Nuclear Research \\
of the Russian Academy of Sciences,\\
60th October Anniversary Prospect, 7a, \\
117312, Moscow, Russia\\
e-mail: berezin@ms2.inr.ac.ru, \\
boyarsk@mech.math.msu.su\\
A.Yu.Neronov \\
Theoretische Physik, \\
Universit\"at M\"unchen,\\
D-80333, M\"unchen, Germany\\
e-mail: neronov@atp.physik.uni-muenchen.de\\}
\date{\today}
\begin{document}
\maketitle
\begin{abstract}
We develop a method for constructing  asymptotic solutions of 
finite~-~difference equations and implement it to a 
relativistic Schroedinger equation
which describes motion of 
a selfgravitating spherically symmetric dust shell. 
Exact mass spectrum of black hole formed due to the collapse of 
the shell is determined from the analysis 
of asymptotic solutions of the equation.
\end{abstract}


\section{Relativistic Schroedinger equation in finite differences}
\label{sec:main}


In this note we develop a method for determining the energy spectrum of 
relativistic Schroedinger equation in finite differences 
 \begin{eqnarray}
 \label{main}
 \Psi (m, \mu, S+i\zeta )+\Psi (m, \mu, S-i\zeta )=\nonumber\\
\left( F_{in}
F_{out}\right)^{-1/2}
\left( F_{in}+F_{out}-M^2/4m^2S\right)
\Psi (m, \mu, S)
 \end{eqnarray}
 
which appear to describe a quantum black hole model with selfgravitating
dust shell introduced in \cite{prd57}. Here parameter $M$ is the
bare mass of the shell, $m=m_{out}$ is the Schwarzchild
mass of the space-time outside the collapsing shell, $\mu =m_{in}/m_{out}$ 
is the ratio of Schwarzchild masses inside and outside the shell, 
$S=R^2/4G^2m^2$ ($G$ is gravitational constant) 
is the dimensionless area of the shell surface and  $F_{in,out}=
1-2Gm_{in,out}/R$ are functions entering the Schwarzchild line element.

Parameter $\left.\zeta=m_{pl}\right/(2m^2)$ where $m_{pl}=\sqrt{\left. c\hbar
\right/ G}$ is Planck mass ($\hbar$ is Planck constant and $c$ is the speed of
light). It becomes small in semiclassical ($\hbar\rightarrow 0$) 
limit. In  this  limit the finite 
shift in the argument of the wave function in (\ref{main}) become small and
one could us an approximate expression
\begin{equation}
  \label{eq:ode}
  \Psi(S\pm i\zeta)\approx\Psi(S)\pm i\zeta\Psi'(S)-\frac{\zeta^2}{2}\Psi''(S)
\end{equation}
Then the equation \ref{main} becomes just a usual Schroedinger of 
nonrelativistic quantum mechanics: 
\begin{equation}
  \label{eq:ode1}
  -\zeta^2 \Psi''+\left\{ 2- \left( F_{in}
F_{out}\right)^{-1/2}
\left( F_{in}+F_{out}-M^2/4m^2S\right)\right\}\Psi =0
\end{equation}
This equation was studied in \cite{prd57,prd58} in details. There was an 
expression for the energy spectrum of bound states found and quasiclassical 
behaviour of the wave function studied.

In this paper we will analyse the solutions of the exact equation 
(\ref{main}).

The shift of the argument in the wave function is along the imaginary line, so
the equation is naturally defined over some complex one dimensional manifold.
 The function
 \begin{equation}
 \label{F}
(F_{in} F_{out})^{1/2}=\sqrt{\left( 1-\frac{1}{\rho}\right)\left(1-\frac{\mu }
{\rho }\right) }=\frac{\sqrt{(\rho -1) (\rho -\mu )}}{\rho}
 \end{equation}
($\rho =\sqrt{S}$) which enter the equation (\ref{main}) 
is a branching function of its argument. So (\ref{main}) is defined over the
Riemannian surface $S_F$ for the function (\ref{F}) which is a two dimensional
sphere glued from the two Riemannian spheres $S_+$ and $S_-$ 
along the sides of the cut made 
along the interval $(\mu, 1)$ of the real line.

There are two points $\rho=\infty$ on the Riemannian surface $S_F$ in $S_+$
and $S_-$ components, which are singular points of equation (\ref{main}). 
They differ by the different chooses of the sign of the function
$(F_{in}F_{out})^{1/2}$.

The natural requirements on the wave function $\Psi$ of a bound state 
is that it must decrease with $\rho\rightarrow\infty$ But in our case we have 
two infinities in $S_+$ and $S_-$ components of $S_F$. As it was argued in
\cite{prd57,prd58,hawk} the two possibilities for  $\rho\rightarrow\infty$ have
the following physical meaning. If one consider maximally extended spherically symmetric solution of Einstein equations in vacuum -- Kruskal manifold 
\cite{misner}, one finds that the radial coordinate $\rho$ take the same 
values on different sides of Einstein-Rosen bridge (in $R+$ and $R_-$ regions 
of Kruskal manifold \cite{prd57}). So $\rho$ could tend to infinity twofold:
either in $R_+$ region or in $R_-$ region. This complicated structure of 
Kruskal manifold is mirrored in complicated structure of configuration space 
of quantum system (the real section of $S_F$) which describe a
selfgravitating dust shell. So the two possibilities for 
$\rho\rightarrow\infty$ in $S_+$ and $S_-$ components of $S_F$ corresponds to
the possibility for $\rho\rightarrow\infty$ in $R_+$ and $R_-$ regions of 
Kruskal space-time. 

But the requirements for the wave function to vanish at infinities are not all
the requirements which a reasonable wave function of a physical state must 
satisfy. Besides if must be one-valued function. 

This requirement is satisfied
automatically if we consider Schroedinger equation over a complex plane
(or over the real line). But the equation (\ref{main}) is defined over 
(the real section of) the Riemannian surface $S_F$. The
two infinities in $S_\pm$ components are both singular points of equation.
So the equation (\ref{main}) is defined in fact over  complex manifold which
is two-dimensional sphere minus two points -- a cylinder. But
a cylinder possess a nontrivial cycle. The monodromy of solution along this
nontrivial cycle is not identity in general. Therefore, the requirement for the
wave function to be one-valued function on the Riemannian surface $S_F$ (and on
its real section) become a nontrivial one. 
In what follows we will find the 
asymptotics of exact solutions of equation (\ref{main}) at both infinities.
The analysis of analytical properties of asymptotics
and the requirement for one-valued solutions to decrease 
at infinities will lead us to defining
of the mass spectrum for the  equation (\ref{main})


\section{Asymptotic solutions}
\label{sec:matrix}


Let us find the asymptotic solutions of this equation at infinities in
$S\pm$ components of $S_F$.

It is convenient to rewrite (\ref{main}) in matrix form. 

If we truncate the Tailor expansion
 \begin{equation}
\Psi (S-i\zeta )=\Psi +\sum_{k=1}^{2N}\frac{(-i\zeta )^k}{k!}\Psi^{(k)}
 \end{equation}
on  $2N$-th item and  expand asymptotically
\begin{eqnarray}
\left( 1-\frac{\mu}{\sqrt{S}}\right)^{-1/2}\left( 1-\frac{1}{\sqrt{S}}\right)^{
-1/2} \approx \nonumber\\ 
 1+\frac{1+\mu }{2}\frac{1}{\sqrt{S}}+
\frac{3\mu^2+2\mu +3 }{8}\frac{1}{S}+\frac{5\mu^3+3\mu^2 +3\mu +5 }{16}
\frac{1}{S^{3/2}}
\end{eqnarray}
in powers of $S$,
we  could rewrite (\ref{main}) as
\begin{eqnarray}
\label{medium}
\Psi^{2N}=-\frac{(2N)!}{(i\zeta)^{2N}}\sum_{k=1}^{N-1}\frac{(i\zeta)^{2k}}
{(2k)!}\Psi^{(2k)}+ \nonumber\\ 
\frac{(2N)!}{(i\zeta)^{2N}}\left( -1\pm 1\pm \frac{(1-\mu)^2
-\left. M^2\right/ m^2}{8S}\right.\pm \nonumber\\ \left. \frac{(1+\mu )\left(
2(1-\mu )^2-\left. M^2\right/ m^2\right)}{16 s^{3/2}}\right)\Psi
\end{eqnarray}
where the upper signs correspond to $S_+$ and the lower ones to $S_-$
components. 
It is convenient to introduce a vector function
\begin{equation}
Y_1=\Psi; \ \ Y_2=\Psi';\ \ ...; \ \ Y_{2N}=\Psi^{2N-1}
\end{equation}
Then (\ref{medium}) takes the form
 \begin{equation}
 \label{matrix}
Y' =AY
\end{equation}
where 
\begin{equation}
A(S)=A_0+\frac{1}{\sqrt{S}}A_1+\frac{1}{S}A_2+\frac{1}{S^{3/2}}A_3
\end{equation}
where $A_0$ has the following form:
\begin{equation}
\label{a0}
A_0=\left(
\begin{array}{ccccccc}
0&1&0&0&...&0&0\\
0&0&1&0&...&0&0\\
.&.&.&...&...&.&.\\
0&0&0&...&...&0&1\\
a_1&0&a_3&0&...&a_{2N-1}&0
\end{array}\right)
\end{equation}
with coefficients
 \begin{eqnarray}
a_1=\frac{(2N)!}{(i\zeta )^{2N}}(-1\pm 1)\nonumber\\
a_{2k}=0\nonumber\\
a_{2k+1}=-\frac{(2N)!}{(2k!)(i\zeta )^{2N-2k}},\ \ k>0
 \end{eqnarray}
Matrices $A_{1,2,3}$ are
\begin{equation}
A_1=0
\end{equation}
and
\begin{equation}
\label{alpha}
A_2=\alpha
\left(
\begin{array}{cccc}
0&0&...&0\\
.&.&...&.\\
0&0&...&0\\
1&0&...&0
\end{array}\right)
;\ \ \alpha =
\pm \frac{(2N)!}{(i\zeta )^{2N}}\frac{\left( 
(1-\mu)^2-\left.M^2\right/ m^2\right) }{8}
\end{equation}
\begin{equation}
\label{beta}
A_3=\beta
\left(
\begin{array}{cccc}
0&0&...&0\\
.&.&...&.\\
0&0&...&0\\
1&0&...&0
\end{array}\right)
;\ \ \beta =\pm\frac{(2N)!}{(i\zeta )^{2N}}\frac{(1+\mu)\left( 
2(1-\mu)^2
-\left.M^2\right/ m^2\right) }{16}
\end{equation}
After the change of variable $S=\rho^2$ equation (\ref{matrix}) takes the form
\begin{equation}
\frac{1}{\rho }Y'_\rho=\tilde AY;\ \ \ \tilde A(\rho )=\tilde A_0+\frac{1}{\rho^2}
\tilde A_2+\frac{1}{\rho^3}\tilde A_3
\end{equation}
with matrices
\begin{equation}
\tilde A_i=2A_i
\end{equation}

If we let $N\rightarrow\infty$ then matrix equation (\ref{matrix}) is 
equivalent to finite~-~difference equation (\ref{main}) in asymptotic regions.
There exist well-developed method for finding asymptotic solutions ($\rho
\rightarrow\infty$) of
matrix equations of (\ref{matrix}) type \cite{vazov}. 

First we need to make a transformation 
\begin{eqnarray}
\label{transformation}
\tilde{\tilde Y}=TY\nonumber\\
\tilde{\tilde A}=T\tilde AT^{-1}\nonumber\\ 
\tilde{\tilde Y}'_\rho =\tilde{\tilde A}\tilde{\tilde Y}
\end{eqnarray}
such that matrix $\tilde A_0$ becomes  diagonal:
\begin{equation}
\tilde{\tilde A}_0 =diag(\lambda_1,...,\lambda_{2N})
\end{equation}
The eigen values $\lambda_1,...,\lambda_{2N}$ could be found by equating
the characteristic polynomial for matrix $\tilde A_0$ to zero:
\begin{equation}
\label{char}
P_{2N}=\left| \tilde A_0-\lambda I\right| =0
\end{equation}
As a result we get an equation
\begin{equation}
P_{2N}=-\frac{(2N)!}{2(i\zeta)^{-2N}}\left(\sum_{k=0}^{N}\frac{1}{(2k)!}
\left( 
\frac{i\zeta\lambda}{2}\right)^{2k}\pm 1\right)=0
\end{equation}
In the limit $N\rightarrow\infty$ this equation is equivalent to the following 
one
\begin{equation}
\label{cosinus}
\cos\left( \frac{\zeta\lambda}{2}\right)=\pm 1
\end{equation}

\begin{figure}

\begin{picture}(0,0)%
\epsfig{file=cosinus.pstex}%
\end{picture}%
\setlength{\unitlength}{3947sp}%
\begingroup\makeatletter\ifx\SetFigFont\undefined%
\gdef\SetFigFont#1#2#3#4#5{%
  \reset@font\fontsize{#1}{#2pt}%
  \fontfamily{#3}\fontseries{#4}\fontshape{#5}%
  \selectfont}%
\fi\endgroup%
\begin{picture}(5194,2168)(303,-4293)
\end{picture}
 \caption{\it Double degeneracy of eigen values in $N\rightarrow\infty$ limit}
\label{fig:cosinus}
\end{figure}

So we see that all the eigen values become double degenerate in the limit
$N\rightarrow\infty$ as it is seen from Fig. \ref{fig:cosinus}.
This means that the matrix $\tilde A_0$ could not be transformed to diagonal
form in this limit, instead it could be transformed to Jordanian form:
\begin{eqnarray}
\label{eqv}
{\cal A}_0=T\tilde{\tilde A_0}T^{-1} =J_1\oplus J_2\oplus ... \oplus J_N;\\
J_i=\lambda_iI+H;\nonumber\\
I=\left(
\begin{array}{cc}
1&0\\
0&1
\end{array}
\right);\ \ \ H=\left(
\begin{array}{cc}
0&1\\
0&0
\end{array}
\right)\nonumber
\end{eqnarray}
It is convenient to rewrite all the matrices in the form of blocks of dimension
$2\times 2$:
\begin{equation}
\label{t1}
T=\left(
\begin{array}{ccc}
T_{11}&...&T_{1N}\\
.&...&.\\
T_{N1}&...&T_{NN}
\end{array}
\right);\ \ \ T_{IK}=\left(
\begin{array}{cc}
t_{11}&t_{12}\\
t_{21}&t_{22}
\end{array}
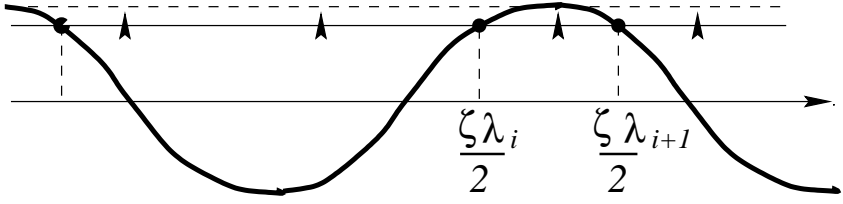
\right)
\end{equation}
and in analogous way
\begin{eqnarray}
\tilde A_0=\left(
\begin{array}{ccc}
A_{11}&...&A_{1N}\\
.&...&.\\
A_{N1}&...&A_{NN}
\end{array}
\right);  \\
A_{II}=\left(
\begin{array}{cc}
0&2\\
0&0
\end{array}
\right);\ \ \ 
A_{I,I+1}=\left(
\begin{array}{cc}
0&0\\
2&0
\end{array}
\right);\ \ \ I=1,...,N-1\nonumber\\ 
A_{NK}=\left(
\begin{array}{cc}
0&0\\
2a_{2K-1}&0
\end{array}
\right);\ K=1,..., N-1; \ \ \ 
A_{NN}=\left(
\begin{array}{cc}
0&2\\
2a_{2N-1}&0
\end{array}
\right)\nonumber
\end{eqnarray}
Then the matrix $T$ could be found from the equation (\ref{eqv}) which
lead to the following system of recurrent equations:
\begin{equation}
A_{II}T_{IJ}+A_{I(I+1)}T_{(I+1)J}=\lambda_{I}T_{IJ}+T_{IJ}H
\end{equation}
Solving this system of equations gives $T$ in the following form
\begin{equation}
\label{t2}
T_{IJ}=\left(
\begin{array}{cc}
\left( \frac{\lambda_J}{2}\right)^{2I-2}t^{(J)}_{11}\ \ \ \ \ &
\left( \frac{\lambda_J}{2}\right)^{2I-2)}t^{(J)}_{12}+
(I-1)\left( \frac{\lambda_J}{2}\right)^{2I-3}t^{(J)}_{11} \\
 & \\
\left( \frac{\lambda_J}{2}\right)^{2I-1}t^{(J)}_{11}\ \ \ \ \  &
\left( \frac{\lambda_J}{2}\right)^{2I-1}t^{(J)}_{12}+
\frac{2I-1}{2}\left( \frac{\lambda_J}{2}\right)^{2(I-1)}t^{(J)}_{11}  \\
\end{array}
\right)
\end{equation}
the coefficients $t^{(J)}_{11}$ and $t^{(J)}_{12}$ are arbitrary and
the form of asymptotic solution of equation (\ref{matrix}) do 
not depend on them. So we will take $t^{(J)}_{11}=t^{(J)}_{12}=1$
for simplicity.

Matrices $\tilde A_{2,3}$ transform under (\ref{transformation}) as follows:
\begin{eqnarray}
\label{L}
\tilde{\tilde A}_2=\alpha 
\left(
\begin{array}{ccc}
L_{11}&...&L_{1N}\\
.&...&.\\
L_{N1}&...&L_{NN}
\end{array}
\right);\\
L_{IK}=\left(
\begin{array}{cc}
\tilde t^{I}_{12}&\tilde t^{I}_{12}\\
\tilde t^{I}_{22}&\tilde t^{I}_{22}
\end{array}
\right)\nonumber
\end{eqnarray}
where $\tilde t^{I}_{ij}$ are elements of the block 
$\left( T^{-1}\right)_{IN}$ of inverse matrix of transformation $T^{-1}$.

The transformed matrix equation (\ref{transformation}) could be reduced to a 
set of $2\times 2$ matrix equations with the help of one more transformation
\begin{equation}
\label{transf1}
\tilde{\tilde Y}=P(\rho)Z
\end{equation}
where 
\begin{equation}
P=\left(
\begin{array}{cccc}
0&P_{12}&...&P_{1N}\\
P_{21}&0&...&P_{2N}\\
.&.&...&.\\
P_{N1}&P_{N2}&...&0
\end{array}
\right)
\end{equation}
is a matrix of $2\times 2$ blocks such that after the transformation
(\ref{transf1}) the system of equations (\ref{matrix}) is rewritten as
\begin{equation}
\frac{1}{\rho}Z'=BZ
\end{equation}
with matrix
\begin{equation}
\label{eq:b}
B=P^{-1}{\tilde A}P-\frac{1}{\rho}P^{-1}P'
\end{equation}
of block-diagonal form:
\begin{equation}
\label{b}
B(\rho)=\left(
\begin{array}{cccc}
B_{11}&0&...&0\\
0&B_{22}&...&0\\
.&.&...&.\\
0&0&...&B_{NN}
\end{array}
\right)
\end{equation}
Expanding 
\begin{eqnarray}
P=P_0+\frac{1}{\rho}P_1+\frac{1}{\rho^2}P_2+...;\nonumber\\
B=B_0+\frac{1}{\rho}B_1+\frac{1}{\rho^2}B_2+...;
\end{eqnarray}
and solving equation (\ref{eq:b}) in each power of $\left. 1\right/ \rho$ one obtains for $B_{ii}$
\begin{equation}
B_{ii}=\lambda_iI+H+\frac{\alpha}{\rho^2}L_{ii}+\frac{\beta}{\rho^3}L_{ii}
\end{equation}
where $\alpha$ and $\beta$ are defined in (\ref{alpha}) and (\ref{beta}) 
and 
\begin{equation}
\label{LL}
L_{ii}=\left(
  \begin{array}{cc}
  l_{1}^i&l_{1}^i\\
  l_{2}^i&l_2^i
  \end{array}
\right)
\end{equation}
is $2\times 2$ matrix (\ref{L}).

The $2N\times 2N$
matrix system (\ref{matrix}) is equivalent to the set of $2\times 2$ matrix systems of the form 
\begin{equation}
  \label{eq:2x2}
  \frac{1}{\rho}Z'={\cal B}Z
\end{equation}
where ${\cal B}$ is one of the $B_{ii}$. 

Now we will construct solutions of the system of ordinary differential 
equations (\ref{eq:2x2}). Let us take
\begin{equation}
  \label{eq:nilp}
  Z=\exp\left(\frac{\lambda_i\rho^2}{2}\right)\left(\left(
    \begin{array}{cc}
1&0\\
0&\frac{1}{\rho}
    \end{array}
\right)+\left(\frac{\alpha}{\rho^2}+\frac{\beta}{\rho^3}\right) 
\left(
  \begin{array}{cc}
0&0\\
-l_1&-\frac{1}{\rho}l_2
  \end{array}\right)\right)U
\end{equation}
Then differential equation for $U$ will be
\begin{equation}
  U'={\cal D}U
\end{equation}
where
\begin{equation}
  \label{eq:d}
  {\cal D}=\left(
    \begin{array}{cc}
0&1\\
\left(\alpha+\frac{\beta}{\rho^2}\right)l_2&
\left(\alpha+1+\frac{\beta}{\rho}\right)(l_1+l_2)
    \end{array}\right) +...
\end{equation}
The last system of equations could be transformed to diagonal form because the
determinant of matrix ${\cal D}_0=\left. {\cal D}\right|_{\rho=\infty}
\not= 0$. The eigen values of matrix 
$$
{\cal D}_0=\left(
  \begin{array}{cc}
0&1\\
\alpha l_2&0
  \end{array}\right)
$$
are $\nu_{1,2}=\pm\sqrt{\alpha l_2}=\pm\nu$. Therefore taking the 
transformation matrix 
$$
{\cal T}=\left(
  \begin{array}{cc}
\frac{1}{2}&\frac{1}{2\nu}\\
\frac{1}{2}&-\frac{1}{2\nu}
  \end{array}\right)
$$ 
one finds the form of transformed system
\begin{eqnarray}
  \label{eq:final}
W={\cal T}U\nonumber\\
{\cal E}={\cal T}^{-1}{\cal D}{\cal T}\nonumber\\
  W'={\cal E}W
\end{eqnarray}
with
\begin{eqnarray}
  {\cal E}=\nu
\left(
  \begin{array}{cc}
1&0\\
0&-1
  \end{array}\right)\nonumber\\
+\frac{1}{\rho}
\left(
  \begin{array}{cc}
\frac{\beta}{2\nu}l_2+\frac{\alpha +1}{2}(l_1+l_2)&
\frac{\beta}{2\nu}l_2-\frac{\alpha +1}{2}(l_1+l_2)\\
-\frac{\beta}{2\nu}l_2-\frac{\alpha +1}{2}(l_1+l_2)&
-\frac{\beta}{2\nu}l_2+\frac{\alpha +1}{2}(l_1+l_2)
  \end{array}\right)
+\nonumber\\ \frac{\beta(l_1+l_2)}{\rho^2}
\left(
  \begin{array}{cc}
\frac{1}{2}&-\frac{1}{2}\\
-\frac{1}{2}&\frac{1}{2}
  \end{array}\right)+...
\end{eqnarray}
Matrices ${\cal E}_{1,2}$ which stand by $1/\rho$ and $1/\rho^2$ 
correspondingly could be transformed to diagonal form as well after 
one more substitution 
\begin{eqnarray}
\label{fine}
  W={\cal P}Q\nonumber\\
Q'={\cal F}Q
\end{eqnarray}
Expanding ${\cal P}$ and ${\cal F}$ in powers of $1/\rho$ one could find the 
form of matrix ${\cal F}$ solving matrix equation
\begin{equation}
  {\cal F}={\cal P}^{-1}{\cal E}{\cal P}-{\cal P}^{-1}{\cal P}'
\end{equation}
The result is
\begin{equation}
  {\cal F}={\cal F}_0+\frac{1}{\rho}{\cal F}_1+...
\end{equation}
where
\begin{eqnarray}
{\cal F}_0=\left(
  \begin{array}{cc}
\nu&0\\
0&-\nu
  \end{array}\right)\\
{\cal F}_1=\left(
  \begin{array}{cc}
0&\frac{\beta}{2\nu}l_2+\frac{\alpha+1}{2}(l_1+l_2)\\
-\frac{\beta}{2\nu}l_2+\frac{\alpha+1}{2}(l_1+l_2)&0
  \end{array}\right)
\end{eqnarray}

Finally we are able to write down the asymptotic solution of system 
(\ref{fine}):
\begin{eqnarray}
  \label{eq:solution}
  Q_1=\exp\left\{\nu\rho+
\left(\frac{\beta}{2\nu}l_2+\frac{\alpha+1}{2}(l_1+l_2)\right) \ln \rho\right\}
\nonumber\\
 Q_2=\exp\left\{-\nu\rho+
\left(-\frac{\beta}{2\nu}l_2+\frac{\alpha+1}{2}(l_1+l_2)\right) \ln \rho\right\}
\end{eqnarray}
where
\begin{eqnarray}
\nu=\sqrt{\alpha}\sqrt{l_2}\nonumber\\
\alpha=\pm\frac{(2N)!}{4(i\zeta)^{2N}}\left((1-\mu)^2-\frac{M^2}{m^2}\right)
\nonumber\\
\beta=\pm\frac{(2N)!}{8(i\zeta)^{2N}}(1+\mu)
\left((2(1-\mu)^2-\frac{M^2}{m^2}\right)
\end{eqnarray}
Using this expression we could restore the asymptotic solution of initial 
matrix system of differential equations (\ref{matrix}) and therefore 
the asymptotic solution of equation (\ref{main}) could be written in the form 
\begin{equation}
  \label{eq:branch}
  \Psi(\rho)=\rho^a
\Phi_1(\rho)+
\rho^b\Phi_2(\rho)
\end{equation}
where $\Phi_{1,2}(\rho)$ are some one-valued functions and $a,b$ are given
by the expressions
\begin{eqnarray}
  \label{eq:powers}
  a=\left(
-\frac{\beta}{2\nu}l_2+
\frac{\alpha+1}{2}(l_1+l_2)
\right)\nonumber\\
b=\left(\frac{\beta}{2\nu}l_2+\frac{\alpha+1}{2}(l_1+l_2)
\right)
\end{eqnarray}

The coefficients $l_1$ and $l_2$ (see (\ref{L}) and (\ref{LL})) 
are expressed through the elements 
$\tilde t_{ij}$ of the matrix inverse to $T$ ((\ref{t1}) and (\ref{t2})).
 Therefore in order to determine the asymptotic expression for solution of 
(\ref{main}) we just need to calculate the inverse matrix $T^{-1}$.

Matrix $T$ has the form
\begin{equation}
  \label{eq:tt}
T=  \left(
    \begin{array}{ccccc}
1&1&1&...&1\\
\lambda_1&\lambda_+\frac{1}{2}&\lambda_2&...&\lambda_N+\frac{1}{2}\\
\lambda_1^2&\lambda_1^2+\lambda_1&\lambda_2^2&...&\lambda_N^2+\lambda_N\\
.&.&.&...&.\\
\lambda_1^{2N-1}&\lambda_1^{2N-1}+\frac{2N-1}{2}\lambda_1^{2N-2}&\lambda_2^{2N-1}&
...&\lambda_N^{2N-1}+\frac{2N-2}{2}\lambda_N^{2N-2}
    \end{array}\right)
\end{equation}
Its determinant will not change if we subtract odd columns from even ones:
\begin{equation}
\det T= \det T_1=\left|
    \begin{array}{ccccc}
1&0&1&...&0\\
\lambda_1&\frac{1}{2}&\lambda_2&...&\frac{1}{2}\\
\lambda_1^2&\lambda_1&\lambda_2^2&...&\lambda_N\\
.&.&.&...&.\\
\lambda_1^{2N-1}&\frac{2N-1}{2}\lambda_1^{2N-2}&\lambda_2^{2N-1}&
...&\frac{2N-1}{2}\lambda_N^{2N-2}
    \end{array}\right|
\end{equation}
The last expression could be written in the following form
\begin{equation}
  \det T=\frac{1}{2^N}
\left.\partial_{\mu_1}...\partial_{\mu_N}W\right|_{\mu_i=\lambda_i}
\end{equation}
where $W$ is Wandermond determinant
\begin{equation}
  W=\left|
    \begin{array}{ccccc}
1&1&1&...&1\\
\lambda_1&\mu_1&\lambda_2&...&\mu_N\\
.&.&.&...&.\\
\lambda_1^{2N-1}&\mu_1^{2N-1}&\lambda_2^{2N-1}&...&\mu_N^{2N-1}
    \end{array}\right|
\end{equation}
Using the well known rule of calculation of Wandermond determinant we obtain 
for $\det T$
\begin{equation}
  \det T=\frac{1}{2^N}\prod_{i>j}(\lambda_i-\lambda_j)^4
\end{equation}
Implementing the same trick for the calculation of corresponding minors 
$M_{ij}$ we obtain the expression for the elements of inverse matrix $T^{-1}$:
\begin{eqnarray}
  \tilde t_{1N}=
2\left(
\frac{1}{\prod_{k=2}^N(\lambda_k-\lambda_1)^2}-
\sum_j
\frac{1}{\lambda_j-\lambda_1}\prod_{k=2}^N
\frac{1}{(\lambda_k-\lambda_1)^2}
\right)
\nonumber\\
\tilde t_{2N}=
-2\prod_{k=2}^N
\frac{1}{(\lambda_k-\lambda_1)^2}
\end{eqnarray}
Then 
\begin{eqnarray}
  l_2=-2\prod_{k=2}^N\frac{1}{(\lambda_k-\lambda_1)^2}\nonumber\\
  l_1+l_2=-2\sum_{j=2}^N\frac{1}{\lambda_j-\lambda_1}\prod_{k=2}^N\frac{1}
{(\lambda_k-\lambda_1)^2}
\end{eqnarray}

The product and the sum entering last equation could be evaluated using the
characteristic equation (\ref{char}) for matrix $A_0$ (\ref{a0}). Indeed, if 
one write the characteristic polynomial $P_{2N}$ in the form
\begin{equation}
  \label{eq:char1}
  P_{2N}=(\lambda-\lambda_1)...(\lambda-\lambda_{2N})
\end{equation}
which for $N$ large enough become approximately
\begin{equation}
  \label{eq:char2}
  P_{2N}\approx (\lambda-\lambda_1)^2...(\lambda-\lambda_{N})^2
\end{equation}
because all the eigen values of matrix $A_0$ become twice degenerate, as it is 
explained earlier, then
\begin{equation}
  \label{eq:char3}
  \left. P_{2N}''\right|_{\lambda=\lambda_1}
\approx (\lambda_1-\lambda_2)^2...(\lambda_1-\lambda_{N})^2
\end{equation}
exactly the required product. Taking into account expression (\ref{cosinus}) 
for $P_{2N}$ in the limit $N\rightarrow\infty$ one easily finds
\begin{equation}
  \label{eq:pr}
  \prod_{k=2}^{N}(\lambda_k-\lambda_1)^2=\frac{(2N)!}{(i\zeta)^{2N}}\left(
\frac{\zeta^2}{4}\right)\cos\left\{\frac{\zeta\lambda_1}{2}\right\}
\end{equation}
In a similar way 
\begin{equation}
  \label{eq:summa}
  \sum_{k=2}^{N}\frac{1}{\lambda_k-\lambda_1}=
-\frac{
\sum_{k=2}^N\prod_{m=2,m\not= k}^N(\lambda_1-\lambda_m)}
{\prod_{k=2}^N(\lambda_1-\lambda_k)}
=-\sqrt{
\left.\frac{P_{2N}^{(4)}}{P_{2n}^{(2)}}
\right|_{\lambda=\lambda_1}}
\end{equation}
and then
\begin{equation}
 \sum_{k=2}^{N}\frac{1}{\lambda_k-\lambda_1}=\pm\frac{\zeta}{2\sqrt{2}}  
\end{equation}

If one substitute these expressions into (\ref{eq:powers}) one finds that the 
powers of $\rho$ which describe the branching of solution at infinity become
\begin{eqnarray}
  \label{eq:spectr}
  a_{\pm}=-\frac{
2(1-\mu)^2-\left. M^2\right/ m^2}{\zeta\sqrt{2}
\sqrt{(1-\mu)^2-\left. M^2\right/ m^2}}\pm\frac{1}{\zeta2\sqrt{2}}\left(
(1-\mu^2)^2-\left. M^2\right/ m^2\right)\nonumber\\
 b_{\pm}=\frac{2(1-\mu)^2-\left. M^2\right/ m^2}{\zeta\sqrt{2}
\sqrt{(1-\mu)^2-\left. M^2\right/ m^2}}\pm\frac{1}{\zeta2\sqrt{2}}\left(
(1-\mu^2)^2-\left. M^2\right/ m^2\right)
\end{eqnarray}
The plus or minus signs are taken at infinities in $S_+$ and $S_-$ components of $S_F$ correspondingly.


\section{Mass spectrum}
\label{sec:mass}



Having determined the form (\ref{eq:branch}) of asymptotic solutions of 
equation (\ref{main}) at infinities in $S_+$ and $S_-$ components of the 
Riemannian surface $S_F$ we can try to impose the condition of decreasing 
of the wave function at both infinities.
But the asymptotic expression
for the solution (\ref{eq:branch}) contains branching functions if  
powers $a$ and $b$ are not integer. This means that the asymptotic expression 
(\ref{eq:branch}) could not be a good approximation for the one valued wave 
function $\Psi$.
It is known from complex analysis, that if the powers $a$ 
and $b$ of $\rho$ in asymptotic expression (\ref{eq:branch}) for the wave 
function $\Psi$ are not integer numbers, the approximation  (\ref{eq:branch})
is valid only in some sector 
$\phi_1<\mbox{ Arg}\left(\left. 1\right/\rho\right)<\phi_2$ in the 
neighbourhood of infinity point. Different approximations are valid in 
different overlaping sectors. This behaviour of the asymptotic approximation 
is called Stokes phenomenon \cite{vazov}.
But if both $a$ and $b$ are integers, the expression (\ref{eq:branch}) is a 
good approximation for non-branching wave function everywhere in the punctured 
neighbourhoods of infinities.\\ 
Due two the double degeneracy of eigen values of matrix $A_0$
the infinite matrix equation is reduced to the infinite set of 
second order matrix equations.
This second order equations can be analysed in the same way as we do this in 
nonrelativistic quantum mechanics. And we  conclude therefore  that the 
wave function with needed boundary conditions on the both sides of the 
Riemannian surface exists only in the situation 
described above -  when $a_{\pm}$ and $b_{\pm}$ are integers. 

The conditions for $a,b$ to be integers are equivalent to
\begin{eqnarray}
  \label{eq:mass}
  \frac{2(1-\mu)^2-\left. M^2\right/ m^2}{\zeta\sqrt{2}\sqrt{(1-\mu)^2-\left.
M^2\right/ m^2}}={\bf n}\nonumber\\
\frac{1}{\zeta2\sqrt{2}}\left((1-\mu)^2-\left. M^2\right/ m^2\right)={\bf k} 
\end{eqnarray}
where {\bf n} and {\bf k} are integers. These conditions define the mass spectrum for the 
equation (\ref{main}).

The the motion of dust shell in its own gravitational field is parameterised 
by three parameters -- the bare mass of the shell $M$, the Schwarzchild mass
which is measured at ``right'' infinity $m=m_{out}$ and the 
Schwarzchild mass inside
the shell (or the mass at ``left'' infinity) $m_{in}=\mu m$. There are
two equations (\ref{eq:mass}) on these three parameters, which we could solve
for two of them. Therefore, the spectrum of Schwarzchild mass $m$ is 
parameterised by two discrete parameters {\bf n} and {\bf k} and one continuous
(for example $\mu$). But if we restrict ourselves with situations when the 
gravitational  field in which the shell moves is produced only by the shell
itself, that is there is no space-time singularity inside the shell, then
$\mu =0$ and the mass spectrum of such states is discrete. 

The expression (\ref{eq:mass}) for the mass spectrum is valid only when 
 $(1-\mu)^2-\left. M^2\right/ m^2>0$ when the square root expression entering 
(\ref{eq:mass}) is well defined. This restriction corresponds to classical
restriction on the trajectories of bound motion of the shell. If it is 
satisfied, then there exist some maximal radius of expansion of the shell
$\rho (t)<\rho_{max}$. 

Otherwise the shell could escape to infinity in its classical motion. For such
shells first of conditions (\ref{eq:mass}) disappears and the only left is
\begin{equation}
  \label{eq:mass1}
\frac{1}{\zeta2\sqrt{2}}\left((1-\mu)^2-\left. M^2\right/ m^2\right)={\bf k} 
\end{equation}
It is remarkable that for light-like shells ($M=0$) in their own gravitational
field ($\mu=0$) this quantization condition become (we recall that 
$\zeta=\left. m_{pl}^2\right/ 2m^2$)
\begin{equation}
  \label{eq:light}
  m=m_{pl}\sqrt{2{\bf k}}
\end{equation}

\section{Conclusion}
\label{sec:con}

In this technical note we developed a method of constructing asymptotical
solutions (\ref{eq:branch}) of relativistic Schroedinger equation in finite 
differences (\ref{main}), which
describe a self-gravitating spherically symmetric dust shell of matter 
\cite{prd57}. This finite difference equation is defined over 
Riemannian surface $S_F$ (\ref{F}). The configuration space of the 
system is the real section of this surface and has nontrivial geometry.
This complicated geometry of configuration space of quantum system is due 
to complicated geometric structure of classical space-time manifold 
of self-gravitating shell.
Analysis of the behaviour of asymptotic solution on Riemannian surface $S_F$ 
enabled us to write down quantization conditions (\ref{eq:mass}) under which
the one valued solution  of (\ref{main}) with physical boundary conditions 
exists. These quantization conditions
define discrete spectrum for Schwarzchild mass $m$ if the mass inside the 
shell is zero. The main result of this paper is that this spectrum up to 
the numerical factors coincides with the spectrum, found in \cite{prd57}
in large black hole approximation and in \cite{prd58} in quasiclassical
approximation.
 This discrete mass spectrum is of the form
\begin{equation}
  \label{eq:bek}
  m\sim m_{pl}\sqrt{{\bf k}}
\end{equation}
for the light-like shells (shells with bare mass $M=0$). This form of quantum
black hole spectrum was extensively discussed by many authors 
\cite{bek,muk}.

\end{document}